\documentclass{ws-procs9x6-cpt22}
\begin{document}

\newcommand{\refeq}[1]{(\ref{#1})}
\def\etal {{\it et al.}}

\title{Searching for Lorentz-violating Signatures from Astrophysical Photon Observations}

\author{Jun-Jie Wei$^{1,2,3}$}

\address{$^1$Purple Mountain Observatory, Chinese Academy of Sciences,\\
Nanjing 210023, China}

\address{$^2$School of Astronomy and Space Sciences, University of Science and Technology of China, Hefei 230026, China}

\address{$^3$Guangxi Key Laboratory for Relativistic Astrophysics,\\
Nanning 530004, China}


\begin{abstract}
As a basic symmetry of Einstein's theory of special relativity, Lorentz invariance has withstood very strict tests.
But there are still motivations for such tests. Firstly, many theories of quantum gravity suggest violations of
Lorentz invariance at the Planck energy scale. Secondly, even minute deviations from Lorentz symmetry can accumulate
as particle travel across large distances, leading to detectable effects at attainable energies. Thanks to their
long baselines and high-energy emission, astrophysical observations provide sensitive tests of Lorentz invariance
in the photon sector. In this paper, I briefly introduce astrophysical methods that we adopted to search for
Lorentz-violating signatures, including vacuum dispersion and vacuum birefringence.
\end{abstract}

\bodymatter

\section{Introduction}
Although experimental tests of Lorentz invariance violation (LIV) have been performed in a wide range of systems
(see Ref.\ \refcite{datatables} for a compilation of results), there are still motivations for such tests. From
theoretical consideration, establishing a quantum theory that includes gravity is being hailed as the Holy Grail
of modern physics. However, theories of quantum gravity (QG) predict that Lorentz invariance may be violated at the
Planck energy scale $E_{\rm Pl}$. From experimental feasibility, even very tiny deviations from Lorentz symmetry can
become measurable at attainable energies $\ll E_{\rm Pl}$, since Lorentz-violating effects gradually accumulate
over large propagation distances. In brief, both the prospect of relativity violations arising in a grand unified
theory and the feasibility of discovering LIV have attracted physicists to constantly work on experimental searches.

Astrophysical observations involving long baselines are very suitable to search for LIV effects. In vacuum,
the LIV-induced modifications to the photon dispersion relation can produce rich and detectable astrophysical phenomena.
With a slight modification, the speed of light will no longer be independent of frequency and polarization.
If the modified dispersion relation is related to the frequency of photons, then we may observer a frequency-dependent
vacuum dispersion of light. If the modification is related to the circular polarization state of photons, the photons
with right- and left-handed polarization states have different velocities, then we may observer vacuum birefringence.
Vacuum dispersion can be tested by measuring the arrival-time differences of photons with different frequencies
originating from the same astrophysical source (see, e.g., Ref.\ \refcite{Amelino-Camelia}).
Vacuum birefringence results in a frequency-dependent rotation of the polarization vector of linearly polarized light,
which can be tested by astrophysical spectropolarimetric measurements (see, e.g., Refs.\ \refcite{Kostelecky,Laurent}).

In this paper, we present some recent tests of nonbirefringent LIV using spectral-lag transitions of gamma-ray bursts
(GRBs),\cite{Wei2017,Du} and some searches for LIV using polarization measurements of GRBs and blazars.\cite{Wei2019,Zhou}

\section{Vacuum dispersion}\label{aba:sec2}
Vacuum dispersion is a potential LIV signature. The arrival time delay between photons with different energies
($E_{h}>E_{l}$) emitted simultaneously from the source at redshift $z$ can be derived by introducing the LIV terms
in a Taylor series:\cite{Jacob}
\begin{equation}
\Delta t=s_{\pm}\frac{1+n}{2H_{0}}\frac{E_{h}^{n}-E_{l}^{n}}{E_{{\rm QG}, n}^{n}}
\int_{0}^{z}\frac{(1+z')^{n}dz'}{\sqrt{\Omega_{\rm m}(1+z')^{3}+\Omega_{\Lambda}}}\;,
\label{eq:tLIV}
\end{equation}
where $s_{\pm}=\pm1$ is the sign of LIV, corresponding to the ``subluminal'' ($s_{\pm}=+1$) or ``superluminal''
($s_{\pm}=-1$) scenarios. $E_{\rm QG}$ denotes the QG energy scale at which LIV effects become significant, $n=1$
($n=2$) corresponds to the linear (quadratic) energy dependence, and ($H_{0}$, $\Omega_{\rm m}$, $\Omega_{\Lambda}$) are
the cosmological parameters of the standard flat $\Lambda$CDM model.

GRBs are ideal astrophysical phenomena that one can used to perform time-of-flight tests because they are the most distant
transient sources involving a wide range of photon energies. However, a key challenge in such tests is to distinguish an
intrinsic astrophysical time delay at the source from a time delay induced by LIV. We proposed that GRB 160625B, the burst
having an apparent transition from positive to negative spectral lags, provides a good opportunity to disentangle the
intrinsic time delay problem.\cite{Wei2017} Spectral lag, the arrival time difference between high- and low-energy photons,
is conventionally defined to be positive if the high-energy photons precede the low-energy ones. In the subluminal case
($s_{\pm}=+1$), photons with higher energies would arrive at the observer after those with lower ones, implying a negative
spectral lag due to LIV. Assuming the source-intrinsic time lag to have a positive dependence on the photon energy, and
considering the contributions to the spectral lag from both the intrinsic positive lag and LIV-related negative lag,
we derived new limits on linear and quadratic leading-order Lorentz-violating vacuum dispersion by directly fitting
the spectral lag behavior of GRB 160625B. Recently, similar time-of-flight tests were carried out by analyzing the
spectral-lag transition of GRB 190114C.\cite{Du}

\section{Vacuum birefringence}
Vacuum birefringence is another potential LIV signature. For a source at redshift $z$, the LIV-induced rotation angle
of the polarization vector of a linearly polarized wave is\cite{Laurent}
\begin{equation}\label{eq:theta-LIV}
  \Delta\phi(E)\simeq\eta\frac{E^2 }{\hbar E_{\rm Pl}H_{0}}\int_0^z\frac{(1+z')dz'}{\sqrt{\Omega_{\rm m}(1+z')^3+\Omega_{\Lambda}}}\;,
\end{equation}
where $E$ is the energy of the observed light and $\eta$ is a dimensionless parameter characterizing the degree of
LIV effects.

If the rotation angles of photons with different energies differ by more than $\pi/2$ over an energy band
($E_{1}<E<E_{2}$), significant depletion of the initial polarization of the signal is expected. The detection of
high polarization can therefore set upper limits on the birefringent parameter $\eta$.
We gave a detailed calculation on the GRB polarization evolution arising from the birefringent effect, and confirmed that
the initial polarization is not significantly depleted even if the differential rotation angle
$|\Delta\phi(E_{2})-\Delta\phi(E_{1})|$ is as large as $\pi/2$.\cite{Wei2019} Applying our formulate for calculating
LIV-induced polarization evolution to the gamma-ray polarimetric data of 12 GRBs, we improved existing bounds on the
birefringent parameter $\eta$ by factors ranging from 2 to 10.

If all photons in the observed energy range are assumed to be emitted with the same intrinsic polarization angle,
we expect to observe vacuum birefringence as an energy-dependent linear polarization vector.
We tried to search for an energy-dependent change of the linear polarization angle in the spectropolarimetric data of 5 blazars.\cite{Zhou} At the $2\sigma$ confidence level, the absence of the birefringent effect was limited to be in the range of
$-9.63\times10^{-8}<\eta<6.55\times10^{-6}$. As might be expected, optical polarimetric data of blazars pose less
stringent constraints on $\eta$ as compared to gamma-ray polarizations of GRBs.

\section{Summary}
GRBs are promising astrophysical sources for searching for LIV-induced vacuum dispersion and vacuum birefringence.
Future detections of extremely-high-energy emission from GRBs with LHAASO, MAGIC, HAWC, and the future international
Cherenkov Telescope Array could improve the limits on LIV using vacuum dispersion (time-of-flight) tests.
Future X-ray/gamma-ray polarization measurements of GRBs with POLAR-II, TSUBAME, COSI, and GRAPE could also
improve the limits on LIV using vacuum birefringence tests.

\section*{Acknowledgments}
This work is partially supported by the National Natural Science Foundation of China
(grant Nos.~11725314 and 12041306), the Key Research Program of Frontier Sciences (grant No.
ZDBS-LY-7014) of Chinese Academy of Sciences, the Major Science and Technology
Project of Qinghai Province (2019-ZJ-A10), the Natural Science Foundation of
Jiangsu Province (grant No. BK20221562), the China Manned Space Project (CMS-CSST-2021-B11),
and the Guangxi Key Laboratory for Relativistic Astrophysics.


\begin{thebibliography}{xx}

\bibitem{datatables}
{\it Data Tables for Lorentz and CPT Violation,}
V.A.\ Kosteleck\'y and N.\ Russell,
2022 edition,
arXiv:0801.0287v15.

\bibitem{Amelino-Camelia}
G. Amelino-Camelia \etal, Nature {\bf 393}, 763 (1998);
A.A. Abdo \etal, Nature {\bf 462}, 331 (2009);
V.\ Vasileou \etal,
Phys.\ Rev.\ D {\bf 82}, 122001 (2013).

\bibitem{Kostelecky}
V.A.\ Kosteleck{\'y} and M.\ Mewes,
Phys.\ Rev.\ Lett. {\bf 87}, 251304 (2001);
V.A.\ Kosteleck{\'y} and M.\ Mewes,
Phys.\ Rev.\ Lett. {\bf 110}, 201601 (2013).

\bibitem{Laurent}
P.\ Laurent \etal, Phys.\ Rev.\ D {\bf 83}, 121301 (2011);
K.\ Toma \etal, Phys.\ Rev.\ Lett. {\bf 109}, 241104 (2012).

\bibitem{Wei2017}
J.-J.\ Wei \etal, Astrophys.\ J.\ Lett. {\bf 834}, L13 (2017).

\bibitem{Du}
S.-S.\ Du \etal, Astrophys.\ J. {\bf 906}, 8 (2021).

\bibitem{Wei2019}
J.-J.\ Wei, Mon.\ Not.\ Royal Astron.\ Soc. {\bf 485}, 2401 (2019).

\bibitem{Zhou}
Q.-Q.\ Zhou \etal, Galaxies {\bf 9}, 44 (2021).

\bibitem{Jacob}
U.\ Jacob and T.\ Piran,
J.\ Cosmol.\ Astropart.\ Phys. {\bf 01}, 031 (2008).



\end{thebibliography}
\end{document}